\RequirePackage{fix-cm}
\documentclass[smallextended]{svjour3}       
\smartqed  
\usepackage{graphicx}
\usepackage{amssymb}
\begin{document}

\title{The collapse of supertasks
}

\titlerunning{Supertasks}        

\author{Gustavo E. Romero}


\institute{Instituto Argentino de Radioastronom{\'{i}}a (IAR, CCT La Plata, CONICET) \at
              C.C. No. 5, 1894, Villa Elisa, Buenos Aires, Argentina. \\
              Tel.: +54-221-482-4903\\
              Fax: +54-221-425-4909\\
              \email{romero@iar-conicet.gov.ar}
}

\date{Received: date / Accepted: date}

\maketitle

\begin{abstract}
A supertask consists in the performance of an infinite number of actions in a finite time. I show that any attempt to carry out a supertask will produce a divergence of the curvature of spacetime, resulting in the formation of a black hole. I maintain that supertaks, contrarily to a popular view among philosophers, are physically impossible. Supertasks, literally, collapse under their own weight.

\end{abstract}
\vspace{0.3cm}

\begin{quotation}
\begin{flushright}
\hspace{-1cm} Then it is again clear that nothing will remain, but it will be all gone... \\[0.5cm]
{\sl Simplicius.}
\end{flushright}
\end{quotation}  

\section{Introduction}
\label{intro}

In recent years there has been a populous literature on supertasks and the strange consequences they have for classical and even quantum physics. A supertask is an {\sl infinite} sequence of {\sl actions} that can be performed in a {\sl finite} interval of time. The term `supertask' was introduced by Thomson in the 1950s in a famous paper (Thomson 1954). The topic experienced a kind of revival after P\'erez Laraudogoitia proposed a new kind of ``beautiful supertask'' (P\'erez Laraudogoitia 1996). This turned the discussion from the logical possibility to the physical implications of supertasks. Some supertasks might, perhaps, be logically possible. I am not concerned with this issue here. My aim is to show that, contrary to a widespread opinion, no supertask can be accommodated in the real universe. I shall show that this follows from some very elementary features of general relativity: any attempt to implement a supertask will produce a black hole. The formation of the black hole will stop the execution of the would-be supertask, and will kill or destroy the executant. 

Some clarifications are in order first. Although my demonstration will be quite general, I shall present a few examples of supertasks for pedagogical reasons, and I shall divide supertasks in two classes. Then, I shall provide some ontological background required by the subsequent discussion. A technical detail goes to the Appendix.          


\section{Supertasks}
\label{sec:ST}

I shall divide supertasks in two groups: class I and class II supertasks. Every action in supertasks of the first class requires the same amount of energy. Supertasks of class II, on the contrary, demand a decreasing amount of energy in every step implemented. Let us review some examples. 

Supertasks of class I are, for instance, Thomson's famous lamp (Thompson 1954) and P\'erez Laugdogoitia's infinite particle collection (P\'erez Laugdogoitia 1996, see also his 2011 review for additional examples and details). In the first supertask, a lamp is just in one of two states: on and off. At $t_1$ the lamp is off. Then, at $t_2=t_1+\Delta t$ the lamp is on. At $t_3=t_2+\frac{\Delta t}{2}$, it is off. At $t_4=t_3+\frac{\Delta t}{4}$, it is on. And so forth. After a finite time\footnote{Since $$ \frac12+\frac14+\frac18+\frac{1}{16}+\cdots = \sum_{n=0}^\infty \frac12\left({\frac 12}\right)^n = \frac {\frac12}{1-\frac 12} = 1. $$} $2\Delta t$, an infinite number of actions have been performed. Thomson's argued that the state of the lamp at the end of the supertask is indeterminate, implying a contradiction. After Benacerraf (1962) criticisms, it is generally accepted that there is no logical contradiction in the performance of this kind of supertasks. The main point is that a property shared by partial sums of a series does not have to be necessarily shared by the limit to which those partial sums tend. For instance, all partial sums involved in the supertask's infinite series are smaller than 1, but the limit is not; it \emph{is} 1. 

In P\'erez Laugdogoitia's ``beautiful'' supertask we are invited to consider an infinite collection of point masses all of which are stationary. The point masses are all of equal mass $m$ and are placed along a line $AB$ at positions $B$, $AB / 2$, $AB / 4$, $AB / 8$, and so on. The first particle at $B$ is accelerated to a velocity $v$ towards A. According to the laws of Newtonian mechanics, when the first particle collides with the second, it will come to rest and the second particle will inherit its velocity $v$. This process will continue through an infinite number of collisions, and after a time $\Delta t=\overline{AB}/ v$, all collisions will finish since all particles were moving at the same velocity $v$. The supertask has been completed. Nevertheless, no particle will emerge from A, since there is no last particle in the sequence. P\'erez Laugdogoitia argues that since all particles are at rest after the supertask completion, the conservation of energy is violated. Moreover, we are told that because of the laws of Newtonian mechanics are time-reversal-invariant, an infinite system of stationary point masses placed along $AB$ will, at random, spontaneously start colliding with each other, resulting in a particle moving away from $B$ at an arbitrary speed. I shall argue in Section \ref{sec:2} that these effects are just a consequence of the incompleteness of Newtonian physics and are not possible in general relativity.

In both examples presented, the amount of energy exchanged in every step of the implementation of the supertask is the same. In class II supertasks, this amount can be progressively 	
diminished. Here there are two examples.

In order to write all natural numbers in one minute one can draw a line dividing an A4 sheet in two equal parts and write `1' in the upper part. After 1/2 minutes, the lower part of the sheet is divided in two and the number `2' is written in the left part. Then, after 1/4 minutes, the remaining part is divided into two, and `3' is written in the upper part. We continue this way until after a minute we have written the infinite natural numbers. Each number is smaller so the work necessary to write it is also smaller. The size of the numbers can be adjusted in order to avoid the necessity of increasing the speed of writing, eluding in this way kinematic superluminal problems.

The other example is the classical Zeno's dichotomy paradox: Achilles wishes to move from $A$ to $B$. To achieve this he must traverse half the distance from $A$ to $B$. To get from the midpoint of $AB$ to $B$ Achilles must traverse half this distance, and so on and so forth. Never mind how many times he performs one of these ``traversing'' tasks there is another one left for him to do before he arrives at $B$. Notice that this will be only a supertask if Achilles stops at the end of each step and starts again to move. Otherwise, if the division is only made in our analysis of motion, there is just one task performed in the physical world: moving from $A$ to $B$. If, instead, there are infinite posts between $A$ and $B$, Achilles must complete infinite actions in a finite time. The energy required to run through each interval can be considered proportional to the distance traversed. Hence, this task can classify as of class II.

If we want now to proceed with a discussion of the physical possibility of class I and II supertasks, we need to give a precise meaning to the word `action' that appears in the definition of supertasks given in the Introduction. To do that, I shall first to provide some basic ontological concepts.

\section{Ontological background}
\label{sec:1}

An individual endowed with properties makes up a thing $X$ (I follow Bunge's ontology). Things can be represented by pairs of the form:
\[
	X=<x,P(x)>,
\]	
where $P(x)$ is the collection of properties of the individual $x$. A thing is an individual with material properties, {\em i.e.} properties that can change in some respect. The most general property of a thing is energy. It is the property of changeability (Bunge 1977). We say that a thing is {\sl material} if it can change. Concepts do not change, since they do not have energy. What changes is the brain that produces concepts.  

The {\em state} of a thing $X$ is represented  by a set of functions $S(X)$ from a 
domain of reference $M$ (a set that can be numerable or non-denumerable\footnote{In most physically interesting cases $M$ is a space-time continuum. See Bunge (1977) and Romero (2013a) for details.}) to the set of properties ${\cal P}_{X}$. Every function in $S(X)$ represents a property in ${\cal P}_{X}$. The set
of the {\sl physically accessible} states of a thing $X$ is the {\em lawful state space} of
$X$: $S_{\rm L}(X)$. The state of a thing is represented by a point in
$S_{\rm L}(X)$. A change of a thing is represented by an ordered pair of states. 

A {\em legal statement} is a restriction  upon the state
functions of a given class of things. A {\em natural law}
is a property of a class of material things represented by an empirically corroborated legal statement.

The {\em ontological history} $h(X)$ of a thing $X$ is a subset of
$S_{\rm L}(X)$ defined by
\[
	h(X) = \{ \langle t, F(t) \rangle | t \in M\} 
\] 
where $t$ is a real parameter on $M$,
and $F$ are the functions that represent the properties 
of $X$.

If a thing is affected ({\em acted upon}) by other things we define:\\
  
$h(Y/X)$: ``history of the thing $Y$ in presence of the thing $X$''.\\

Let $h(X)$ and $h(Y)$ be the histories of the things $X$ and $Y$, respectively.\\
Then 
\[
	h(Y/X) = \{ \langle t, H(t) \rangle |\: t \in M\}, 
\] 
where $ H\neq F $ is the total state function of $Y$ as affected by the existence of $X$, and $F$ is the total state function of $X$ in the absence of $Y$. The history of $Y$ in presence of $X$ is different from the history of $Y$ without $X$.\\

We now introduce the notion of {\sl action}:\\

$ X \vartriangleright Y $: ``$X$ acts on $Y$''
\[
 X \vartriangleright Y \stackrel{\rm def}{=} h(Y/X)\neq h(Y) 
\] 

 An action modifies the state of a thing, and hence its history.

 An {\sl event} is the result of an action, and it is represented by an ordered pair of states:
\[ (s_1, s_2 ) \in E_{\rm L}(X) = S_{\rm L}(X) \times S_{\rm L}(X) \]

The space $E_{\rm L}(X)$ is called
the {\em event space} of $X$.

Any action that causes a change of state implies an exchange of energy. Any change in the energy density of a physical system implies a change in the geometry of the manifold that represents space-time, which is the ontological composition of all events (changes). See Perez Bergliaffa et al. (1998) and Romero (2013a) for an axiomatic theory of spacetime as a system of physical events.

\section{The collapse of class I supertasks}
\label{sec:2}
 
From the considerations outlined above it follows that there is no action without energy. In general relativity energy and momentum of physical systems determine through Einstein's field equations (EFEs) the metric of spacetime, and hence its curvature:
\begin{equation}
R_{\mu\nu}-\frac{1}{2}R g_{\mu\nu}=\frac{8\pi G}{c^{4}}T_{\mu\nu}.
\end{equation}
Here, $T_{\mu\nu}$ is the energy-momentum tensor, $R_{\mu\nu}$ and $R$ are the Ricci tensor and the Ricci scalar, related to spacetime curvature\footnote{Both the Ricci tensor and the Ricci scalar are zero in a flat spacetime since they are formed with the second derivatives of the metric.}, and $g_{\mu\nu}$ is the spacetime metric. The relation between coordinates ${x^{\mu}}$ and the spacetime interval $ds$ between events arbitrarily proximate is $ds^{2}=g_{\mu\nu}dx^{\mu}dx^{\nu}$.

By definition, any supertask is performed in a finite region of spacetime. We can express the volume of this region in an invariant way:
\begin{equation}
V_{\rm ST}=\int_{\rm ST} \sqrt{-g} dx^{4},
\end{equation}
where $g\equiv det[g_{\mu\nu}]$ is the metric determinant and ST represents the spacetime region where the supertask is carry out. The execution of a class I supertask requires that each action must take place in an ever smaller region of spacetime, i.e. $V_{ST}\rightarrow 0$ as $n\rightarrow \infty$, with $n$ the number of steps in the supertask. At the same time, the energy used in each step, $\Delta E$, remains constant. Hence, the energy density $\rho\rightarrow \infty$ as $n\rightarrow \infty$. The divergence of the energy density makes the $tt$ component of $T_{\mu\nu}$ to diverge, and through EFEs, the curvature diverges as well. The process, since the volume is decreasing, is identical to that of gravitational collapse: the singularity (i.e. the divergence of the curvature) will be covered by an event horizon according to the so-called Cosmic Censorship Conjecture (CCC, see, e.g., Hawking \& Ellis 1973). In other words, the attempt to complete the supertask will lead to the formation of a black hole\footnote{Some peculiar choices for $T_{\mu\nu}$ might lead to a divergence of the curvature without the occurrence of an event horizon, in violation of the CCC (e.g. Joshi 1993). Whether such configurations of energy and momentum can be obtained in Nature is highly controversial. In any case, the execution of the supertask is terminated not by the event horizon, but by tidal forces associated with the increase of the curvature which necessarily follows from the rise of the energy density. }.  

A black hole is a region of spacetime from where no time-like\footnote{A time-like curve is a curve that can be followed by a massive system.} or null\footnote{A null curve is a curve that can be followed by photons or other bosons moving at the speed of light.} curve can emerge and reach the conformal future infinity\footnote{A region to which can evolve causal curves from a region located in the past of the black hole.}. In the interior of the black hole the tidal forces increase without bounds, destroying any agent responsible for the supertask. Strictly speaking, the supertask can never be finished.  

\section{The collapse of quantum supertasks}
\label{sec:3}

A supertask enforcer can take refuge in class II supertaks, hoping that at some point quantum effects might avoid the spacetime collapse. This hope, like most, is futile.

Every supertask of class II will eventually reach a quantum regime. The amount of energy involved in the performance of the supertask will be restricted by Heisenberg's inequality for energy\footnote{See the Appendix.}:
\begin{equation}
\frac{\Delta A}{\left|d\left\langle A\right\rangle/dt\right|}\; \Delta E \geq \frac{1}{2}\hbar,
\end{equation}  
or
\begin{equation}
\tau_{A}\; \Delta E \geq \frac{1}{2}\hbar, \label{HeisenbergE}
\end{equation}  
where
\begin{equation}
\tau_{A}=\frac{\Delta A}{\left|d\left\langle A\right\rangle/dt\right|}.
\end{equation}  
In these equations, $A$ is the quantum operator representing the property of the system that is affected by the supertask. The mean value $\left\langle A \right\rangle$ is:
\begin{equation}
\left\langle A\right\rangle= \left\langle \psi \left|A\right|\psi\right\rangle,
\end{equation}
with $\psi$ the vector of the Hilbert space representing  the dynamical state of the system at a given time $t$. The mean root-square deviations of $A$ and the Hamiltonian $H$ are 
$\Delta A=(A-\left\langle A\right\rangle) |\psi \rangle$ and 
$\Delta E=(H-\left\langle H\right\rangle) |\psi \rangle$, respectively.

As a supertask proceeds, $\tau_{A}\rightarrow 0$ and consequently, from (\ref{HeisenbergE}), $\Delta E \rightarrow \infty$. These increasingly stronger energy fluctuations will occur within a region of size $l_A\sim \tau_A c$, which is the largest region occupied by a coherent quantum supertask. Then, the formation of a black hole is unavoidable. We can estimate the magnitude of the energy fluctuations necessary for the collapse. From equation (\ref{HeisenbergE}), we get:
\begin{equation}
l_{A}\; \Delta E \geq \frac{1}{2}\hbar c.
\end{equation} 
To trigger black hole formation, the length $l_{A}$ must equal the Schwarzschild radius of the equivalent mass corresponding to the energy $\Delta E$: $R_{\rm S}=2GM/c^{2}=2G \Delta E/ c^{4}$. Then,
\begin{equation}
\Delta E ^{2} \approx \frac{1}{4}\frac{\hbar}{G} c^{5},
\end{equation} 
which implies:
\begin{equation}
\Delta E \approx 10^{15}\;\;\; {\rm erg}.
\end{equation} 
The quantum black hole will form when the timescale of the supertask be $\tau_{A}\approx 4.8 \times 10^{-43}$ s. This is an order of magnitude above the Planck time.\footnote{The Planck time is $t_{\rm P} \equiv \sqrt{\frac{\hbar G}{c^5}}\approx 5.39106(32) \times 10^{-44}$ s.} 

 Black hole formation will abort the implementation of any class II supertask even before reaching the domain of quantum gravity. 

\section{Epilogue}
\label{sec:6}

I have shown that any attempt to carried out a supertask, in the long run, will lead to such an increase of the energy density in a finite region of spacetime that the formation of a black hole is unavoidable. This will avert the completion of the supertask, even if it is conceivable to finish it in a Newtonian spacetime. 

The friends of supertasks can claim that, at least in the case of quantum supertasks, quantum gravity effects might prevent the collapse allowing the supertask to be finished. This hope, I think, is vain. The only thing that quantum gravity might preclude is the spacetime singularity inside the black hole, not the black hole itself. Singularities are pathological manifestations of the incompleteness of general relativity (see Romero 2013b), and not a feature of the real world. Hence, any good theory of quantum gravity should be free of spacetime singularities. The event horizons that characterize black holes, however, occur far away from any singularity. Spacetime is well-behaved in the horizon, and general relativity provides an adequate description of a black hole, except from its innermost region. Black holes are predicted by any gravitational theory that is consistent with the current knowledge of the universe. Tidal forces, on the other hand, can destroy any physical system capable of performing any task long before the system reaches the region where the description provided by general relativity fails. Theories of quantum gravity cannot give any harbourage to supertasks. 

Should we conclude that supertasks are, at most, a logical game, with no use at all in physical science? I think not. Supertasks have proved to be a powerful tool to probe the limitations of classical Newtonian mechanics, and even quantum mechanics on flat spacetime. The lack of back-reaction of spacetime to the performance of supertasks inherent to classical and special-relativistic mechanics leads to a number of results that point out the weakness of these theories. Supertasks can produce Cauchy horizons in Newtonian spacetimes, and massive violations of energy and momentum conservation laws. A better understanding of some of our most cherished theories, of their frailty and bugs, is crucial to guide us towards the formulation of more comprehensive theories of reality. Here, perhaps, supertasks might finally find their place in the world.

\section*{Appendix: Quantum indetermination theorems}
\label{Appendix}

The relation (\ref{HeisenbergE}) given above is different from the usual inequality presented in many textbooks:
\begin{equation}
	\Delta E \Delta t \geq \frac{\hbar}{2}.
\end{equation}

This inequality cannot be derived from the framework of quantum mechanics, since, as noted by Bunge (1967) and Perez Bergliaffa et al. (1993) among others, there is no time operator in the theory. The inequalities between properties represented by non-commuting operators are derived using the Schwarz inequality (Weyl 1928), i.e. if $\hat{A}$ and $\hat{B}$ are two operators representing the quantum properties $A$ and $B$ such that
\begin{equation}
	[\hat{A}, \; \hat{B}]=i\hbar
\end{equation}
then, through $(\Delta A)^{2}(\Delta B)^{2}\geq \frac{1}{2}\left|\hat{A}\hat{B}\right|$, we get
\begin{equation}
	\Delta A \Delta B \geq\frac{1}{2} \hbar,
\end{equation}
where $\Delta$ indicates the root-mean-square deviation of the corresponding operators.
In particular, since the position and momentum operators satisfy $[\hat{q_{i}}, \; \hat{p_{j}}]=i \hbar \delta_{ij}$, the usual relation $\Delta q_{r} \Delta p_{r}\geq \hbar/2$ obtains. 

This procedure fails when applied to the Hamiltonian that represents the energy of the system and time, since time is a parameter. In 1945, Mandelshtam and Tamm obtained the correct form for an energy-time relation, where the time involved was the timescale of an evolving operator. This relation is given by inequality (\ref{HeisenbergE}).



\begin{thebibliography}{9}

\bibitem{Bena1962} P. Benacerraf (1962). Tasks, super-tasks, and modern Eleatics, {\itshape Journal of Philosophy} {\bf LIX}, 765-784

\bibitem{Bunge1967}
M. Bunge (1967). {\itshape Foundations of Physics}. New York: Springer-Verlag.

\bibitem{Bunge1977}
M. Bunge (1977). {\itshape Ontology I: The Furniture of the World}. Dordrecht: Kluwer.



\bibitem{H-E} 
S.W. Hawking, and G.F.R. Ellis (1973). {\itshape The Large Scale Structure of Space-Time}. Cambridge: Cambridge University Press. 

\bibitem{Joshi} 
P.S. Joshi (1993). {\itshape Global Aspects in Gravitation and Cosmology}.  Oxford Clarendon Press.

\bibitem{Tamm}
 L. I. Mandelshtam, and I. E. Tamm (1945). The uncertainty relation between energy and time in nonrelativistic quantum mechanics, {\itshape Journal of Physics} {\bf IX}, 249-254.

\bibitem{Bergliaffa2}
S.E. Perez Bergliaffa, G. E. Romero,  and H. Vucetich (1993). Axiomatic foundations of non-relativistic quantum mechanics: A realistic approach, {\itshape  Int. J. Theor. Phys.} \textbf{32},  1507-1522.


\bibitem{Bergliaffa1}
S.E. Perez Bergliaffa, G. E. Romero,  and H. Vucetich (1998). Toward an axiomatic pregeometry of space-time, {\itshape  Int. J. Theor. Phys.} \textbf{37},  2281-2298.

\bibitem{PL1}
J. P\'erez Laraudogoitia (1996). A beautiful supertask, {\itshape Mind}, {\bf 105}, 81-83.


\bibitem{PL2}
J. P\'erez Laraudogoitia, (Spring 2011 Edition), Supertasks, {\itshape The Stanford Encyclopedia of Philosophy}, Edward N. Zalta (ed.),\\ URL = http$:$//plato.stanford.edu/archives/spr2011/entries/spacetime-supertasks/.

\bibitem{T}
J. Thomson, (1954). Tasks and Super-Tasks, {\itshape Analysis}, \textbf{XV}, 1�13.



\bibitem{Romero2013a}
G.E. Romero (2013a), From change to spacetime: an Eleatic journey, {\itshape Foundations of Science} \textbf{18}, 139-148.

\bibitem{Romero2013b}
G.E. Romero (2013b), Adversus Singularitates: The ontology of space�time singularities, {\itshape Foundations of Science} \textbf{18}, 297-306.

\bibitem{Weyl1928}
H. Weyl (1928). \emph{Gruppentheorie Und Quantenmechanik}. Leipzig: Hirzel

\end{thebibliography}

\bibliographystyle{aipproc}   


\newpage

\section*{Gustavo E. Romero} Full Professor of Relativistic Astrophysics at the University of La Plata and Chief Researcher of the National Research Council of Argentina. A former President of the Argentine Astronomical Society, he has published more than 300 papers on astrophysics, gravitation, and the foundation of physics, and 9 books (including {\sl Introduction to Black Hole Astrophysics}, with G.S. Vila, Springer, 2013). His main current interest is on black hole physics and ontological problems of space-time theories.

\end{document}